\documentclass[epj,twocolumn]{webofc}
\usepackage[varg]{txfonts}   
\woctitle{MPGD2015}
\begin{document}
\title{Microbulk Micromegas for the search of 0$\nu\beta\beta$ of 136Xe in the PandaX-III experiment}
%
%

\author{J. Galan\inst{1,2}\fnsep\thanks{\email{javier.galan.lacarra@cern.ch}} on behalf of the PANDAX-III Collaboration }

\institute{Shanghai Jiao Tong University, Shanghai, China.
\and Grupo de Física Nuclear y Astropartículas, University of Zaragoza, Zaragoza, Spain.  }

\abstract{%
    The search for the neutrinoless double beta decay (0$\nu\beta\beta$) is one of the most important quests nowadays in neutrino physics. Among the different techniques used, high pressure xenon (HPXe) gas time projection chambers (TPC) stand out because they allow to image the topology of the 0$\nu\beta\beta$ event (one straggling track ending in two blobs), and use it to discriminate signal from background events. Recent results with microbulk Micromegas in Xe + trimethylamine (TMA) mixtures show high promise in terms of gain, stability of operation, and energy resolution at high pressures (up to 10 bar). The addition of TMA at levels of ~1\% reduces electron diffusion in up to a factor of 10 with respect pure Xe, improving the quality of the topological pattern, and therefore the discrimination capability. Moreover microbulk Micromegas have very low levels of intrinsic radioactivity. All these results show that a Micromegas-read High Pressure Xenon TPC (HPXe-TPC) can be a competitive technique in the search for 0$\nu\beta\beta$. The recently proposed PandaX-III experiment, based on these results, aims at building a large TPC of 200 kg of enriched Xe, to be located at Jinping Underground laboratory in China. In this talk the main features of the experiment will be presented, with an emphasis on the design and tests of the microbulk readout, as well as the status of the project and first results of the prototyping phase.  }
\maketitle
\section{Introduction}
\label{intro}
Depending on their mathematical description, fermions can be Majorana or Dirac.  Majorana fermions have been an element of quantum field theory since the very beginning. Whether fermions behave as a Majorana particle is still an open question in the Standard Model of particles. The Majorana description implies that a particle is its own antiparticle. Thus, the only Majorana candidate, between the known elementary particles, is the neutrino, due to its intrinsic neutrality. Today, the most sensitive probe for Majorana neutrinos is a nuclear process known as neutrinoless double-beta decay (0$\nu\beta\beta$), whereby a nucleus decays by emitting two electrons and nothing else~\cite{0nudbd}. A double beta decay occurs when a nucleus is energetically or spin forbidden to decay through single beta decay. A related decay process, known as two-neutrino double beta decay (2$\nu\beta\beta$), was only first observed in 1986 on $^{82}$Se, due to the long lifetimes of double beta decay isotopes. The most precise measurement on the half-life of the 2$\nu\beta\beta$ process has been obtained using $^{136}$Xe~\cite{2nudbdHL}. The half-life of the $0\nu\beta\beta$ process is related to the effective Majorana neutrino mass, and it can be distinguished from the $2\nu\beta\beta$ because the 2 resulting electrons carry the total available Q-value energy, while is a continuum for the latter. The observation of a $0\nu\beta\beta$ process would prove the Majorana nature of fermions, demonstrate lepton number violation and measure the neutrino mass scale~\cite{nReview}.

A successful experiment willing to observe the $0\nu\beta\beta$ decay process needs to guarantee an excellent energy resolution in order to distinguish the Q$_{\beta\beta}$ events produced by this process from the continuum produced by the 2$\nu\beta\beta$ decays, and discriminate other background events expected in the Region Of Interest (ROI) around the Q$_{\beta\beta}$-value. Recent sensitive searches include experiments using $^{76}$Ge (Gerda~\cite{Gerda}) and $^{136}$Xe (EXO-200~\cite{EXO200}). In particular Gerda and EXO200 Collaborations provide the best sensitivities to the half-life of the 0$\nu\beta\beta$ decay, setting $T_{1/2}^{0\nu} \gtrsim 2.1\cdot10^{25}$\,yr and  $T_{1/2}^{0\nu} \gtrsim 1.1\cdot10^{25}$\,yr at 90\% CL., respectively. Next generation experiments searching for 0$\nu\beta\beta$ decay need to increase the mass of the active nuclei, proving scalability towards the tonne scale. The final sensitivity reachable will strongly depend on the background level of the detector, assuring a few counts per year could be observed in the ROI.

The key advantage of a HPXe-TPC resides on the capability to image the tracks of the two emitted electrons in a 0$\nu\beta\beta$ event, reducing the effect of the background on the ROI, defined around $Q_{\beta\beta} = 2457.83$\,keV~\cite{Qbb} for $^{136}$Xe. Natural xenon contains already an 8\% of $^{136}$Xe, and enrichment to 90\% is well stablished by centrifugation methods. Therefore, reaching a tonne scale enriched isotope is cost-effective. A gaseous or liquid xenon TPC allows purification through recirculation and filtering methods, reducing the impurities of the source. The longer tracks and lower diffusion of these events in a gaseous medium will provide enhanced background rejection capabilities when using a gaseous TPC. Another advantage of using a gaseous TPC is the lowest intrinsic energy resolution, which remains constant up to pressures about 50\,bar~\cite{XenonGas}. The construction of a high pressure TPC containing the same amount of mass that a liquid TPC represents a major challenge due to the higher active volume required.

\section{The PandaX-III experiment}\label{pandaXIII}

The current top priority of the PandaX-III is to construct a 4\,m$^3$ low-radioactivity xenon gaseous TPC, working at 10\,bar (equivalent to $\sim$200 kg of xenon gas). Special care is being taken in the design and selection of materials to reach a background level of 10$^{-2}$-10$^{-3}$\,keV$^{-1}$kg$^{-1}$yr$^{-1}$ (before applying topological cuts). A micropatterned charge readout will add the capability of observing real time events with tracking technology. The spatial resolution of this readout should guarantee the capability to differentiate the topology of a 0$\nu\beta\beta$ event from other background events. This unique feature of our detector will allow to reduce even further the intrinsic background level. Preliminary studies on the readout optimization with that purpose are already being carried out.

\begin{figure}[!ht]
\centering
\includegraphics[width=8.5cm,clip]{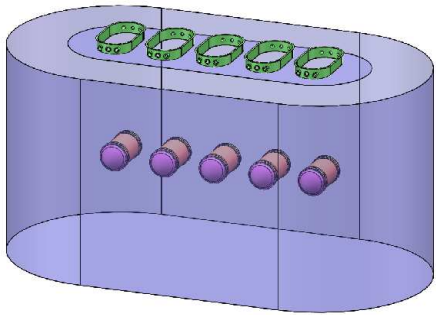}
\caption{Conceptual design of the water tank to be installed at CJPL-II, where the 200kg-TPC modules will be hosted.}
\label{fig:waterTank}       
\end{figure}

PandaX-III will be installed at the China Jinping Underground Laboratory~\cite{Jinping}. It will be placed at the new laboratory upgrade (CJPL-II) which is under construction, to be finished by 2016. The Jinping lab is today the deepest underground laboratory with a depth of 6500\,wme and a cosmic flux below 1\,$\mu$/week/m$^2$. The experiment has already space allocated at one of the eight experimental halls. The size of the PandaX hall will be resized to 14\,m tall, 15\,m wide and 65\,m length in order to fit the experiment needs. A large water tank providing at least 5\,m shielding in each direction is being designed, this tank will be able to host 5 TPCs (200\,kg $^{136}$Xe) modules dedicated to the 0$\nu\beta\beta$ decay search (see Figure~\ref{fig:waterTank}). A purification system is being developed to reduce the $^{238}$U and $^{232}$Th contamination, and to minimize the amount of $^{222}$Rn diluted on water.

\begin{figure*}
    \centering
    \includegraphics[width=14cm,clip]{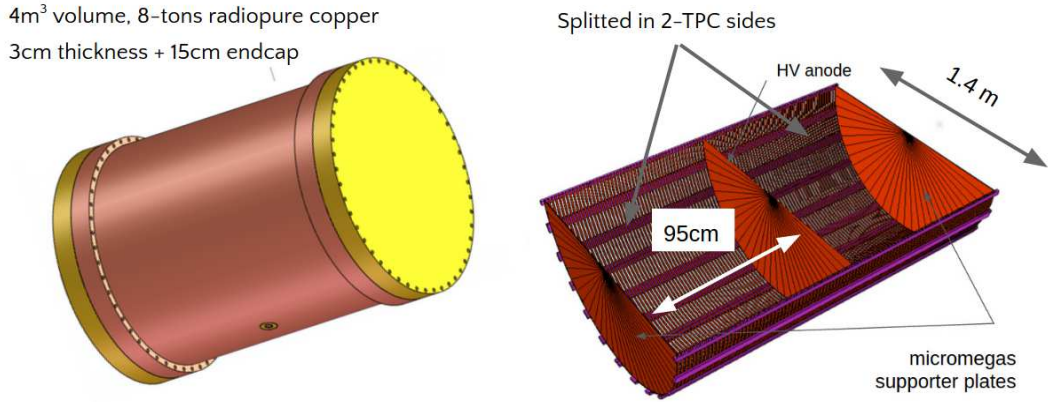}
    \caption{On the left, the high pressure vessel designed to contain 200-kg of xenon at 10\,bar, endcap flanges are highlighted in yellow. On the right, a preliminary design of the field cage to be installed inside the vessel. The field cage consists of equidistant copper rings electrically connected through resistors. The ring structure is supported by several low radiopurity teflon bars (pink color). One high voltage cathode in the middle splits the chamber in two independent TPCs, being readout by two anode planes placed at each side of the vessel. }
    \label{fig:vessel}   
\end{figure*}

A first module is expected to be constructed and installed underground by 2017. A 3\,cm thick radiopure copper vessel will be built, the endcaps of the cylindrical vessel will be 15\,cm thick in order to protect the active volume from the intrinsic radioactivity of the electronics that would be installed just behind. Studies on radiopurity of electronics are being carried out to minimize its effect on the final background level of the detector. A field cage made of copper rings and teflon is being designed to keep a homogeneous field of 1\,kV/cm. The vessel will be splitted on two independent TPC volumes by means of a high voltage central cathode (see Figure~\ref{fig:vessel}).


\section{Micromegas for 0$\nu\beta\beta$ searches}\label{mMDBD}

Micromegas is a micropatterned readout for gaseous TPCs with charge amplification~\cite{micromegas}. The main advantages of this technology for rare event searches resides in its good energy resolution, good spatial resolution providing tracking capability, and the radiopurity of the materials that these readouts are made of. In particular, the microbulk micromegas technology~\cite{microbulk} is specially interesting from the radiopure point of view given its low mass budget and the intrinsic radiopurity of the materials used for its construction (kapton and copper). This technology provides also improved energy resolution thanks to the uniformity of the typically 50\,$\mu$m amplification gap that is made out of a both sides coated 5\,$\mu$m-copper kapton foil.

\subsection{Previous work}\label{mMWork}
Recent progress on micromegas developments, in the framework of the T-REX\footnote{http://gifna.unizar.es/trex} ERC funded project, shows promise on using this technology for the search of 0$\nu\beta\beta$ decay searches~\cite{mMsDBD00}. Radiopurity measurements have shown its low intrinsic radioactivity~\cite{mMsPurity}, proving its potential for using it in rare event search experiments.

It was shown the benefit of using xenon with low amounts of trimethylamine (TMA) to reduce the negative effect of higher pressures on the energy resolution~\cite{mMsXenonTMA}. The achievable energy resolution has been proven to be about 3\%FWHM at the Q$_{\beta\beta}$ for $^{136}$Xe. Although further optimization could improve the energy resolution closer to the intrinsic energy resolution of small micromegas detectors (1.0\% at Q$_{\beta\beta}$). These same studies show also the track reconstruction capabilities using a 8\,mm pixelated micromegas readout~\cite{mMsDBD01}.

An additional advantage of using TMA resides on the lower diffusion of electrons when drifting inside the TPC. A study on 0$\nu\beta\beta$ pattern recognition and background discrimination shows the benefit of using a low diffusion gas mixture on the background rejection, and shows that there is potential for background rejection using pattern recognition algorithms to reduce the background level by two to three orders of magnitude keeping signal efficiency above 40\%~\cite{mMsDBD02}.

\subsection{Micromegas for PandaX-III.}\label{mMPanda}

\begin{figure}[h!]
\centering
\includegraphics[width=7cm,clip]{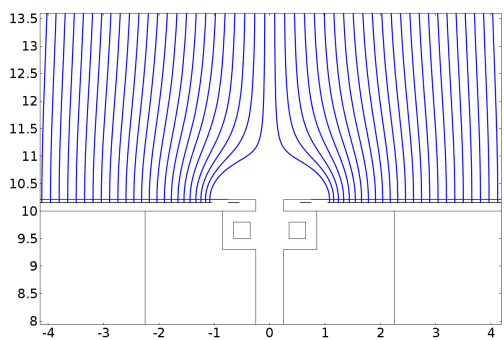}
\caption{Field simulation in the space between 2 micromegas readout modules. The field lines at the drift volume are deflected by using a boundary electrode, or rim electrode, that is connected to a higher voltage than the micromegas mesh. As a result, the electrons will be deviated towards the detector active region avoiding charge losses. Units are expressed in mm. }
\label{fig:rim}       
\end{figure}

The PandaX-III TPC will be readout by two micromegas readout planes of about $\gtrsim$1\,m$^2$. Due to limitations on the construction of large microbulk micromegas readouts each micromegas plane will be composed by a mosaic of micromegas modules of 20$\times$20\,cm$^2$, or Scalable Readout MicroMegas (SR2M). The space between two micromegas modules will be minimized as much as possible, and fixed by mechanical constrains. In order to avoid charge losses in this gap a rim electrode surrounding each micromegas module readout will be placed at a higher voltage to assure the electrons drifting towards this region are deviated and driven towards the active readout region (see Figure~\ref{fig:rim}).

\begin{figure}[h!]
\centering
\includegraphics[width=6cm,clip]{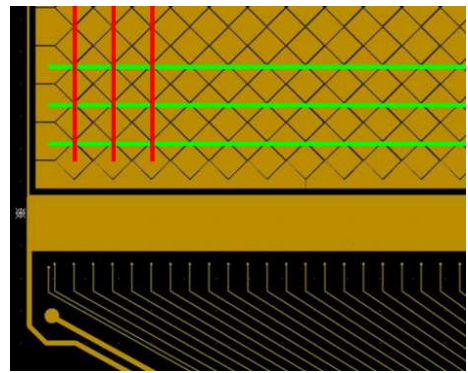}
\caption{A detail of one of the micromegas readouts being manufactured for the PandaX-III experiment. This picture corresponding to a 3\,mm pitch stripped readout. Red vertical lines and green horizontal lines are drawn to show the different pixels interconnectivity, which are readout through the same electronic channel. }
\label{fig:mMreadout}       
\end{figure}

The number of readout channels required for large area coverage is extremely high in the case of pixelized readouts, about 10k channels for each micromegas module for a granularity of the order of $\sim$mm. The complexity of such a design, together with the additional electronic instrumentation (acquisition cards and power supplies) that would have to be installed close to the vessel makes this approach unviable. The electronic setup should be minimized in order to reduce the effect on the final background level of the detector. This mimimization of electronics is first carried out reducing the number of channels to few hundred per micromegas module, by using an interconnected pixels readout scheme (see Figure~\ref{fig:mMreadout}). New studies are being carried out in order to quantify the possible impact of the new readout topology on pattern recognition, however this effect might be mitigated by the higher granularity that can be achieved by using this stripped scheme. 

Actually, three different micromegas module prototypes, with different pitch size (1\,mm, 2\,mm and 3\,mm), are being constructed for testing. The characterization of these prototypes with different calibration sources will be used to determine the effect on the micromegas granularity on the capability of background rejection. The data acquired with these modules will be used to validate the performance of the micromegas modules, the event reconstruction capability, and to compare with full Montecarlo simulation of the events. A large volume prototype (equivalent to about $\sim$10\,kg of xenon at 10\,bar) is being constructed at SJTU, and designed to host 7 micromegas modules. This prototype will be used for testing the different technical issues, as field cage, feedthroughs, study mechanical constrains and test the mounting of micromegas modules. At the same time, the data acquired with this TPC will allow to determine the future performance of the final detector.

\begin{figure}[h!]
    \centering
    \includegraphics[width=8cm,clip]{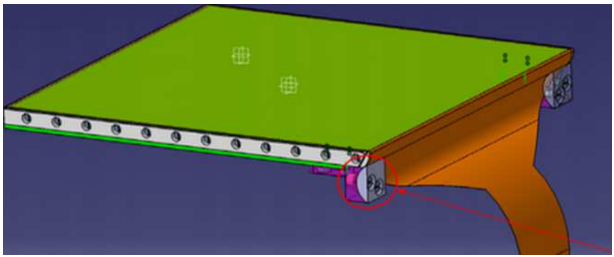}
    \caption{Design of the support where micromegas readout will be glued. The micromegas foil is built with a long extension where readout channels are driven outside the detector without the need of using a connector close to it.  }
    \label{fig:support}   
\end{figure}

The micromegas will be installed in a radiopure copper support specially designed to minimize the gap between two micromegas modules. The micromegas readout, which is built out of a kapton-copper foil, will be extended in such a way that the readout channel signal can be extracted far away the module, avoiding to place a connector (which are typically made of non-radiopure materials) close to the detector region (see Figure~\ref{fig:support}). The length of this extension will be long enough so that it can be extracted directly from the copper vessel. The connector and electronics will be placed behind the copper vessel flange (of at least 15\,cm thickness) minimizing the effect these components could have in the background level of the detector.

\section{Conclusions}
The PandaX-III experiment is preparing to built a first 200\,kg HPXe TPC module to prove the scaling capability of this technology reaching competitive background levels, aiming to reach levels of 10$^{-4}$-10$^{-5}$\,keV$^{-1}$kg$^{-1}$yr$^{-1}$ by exploiting the good discrimination capabilities provided by gaseous TPC built with micromegas technology. This experimental program would allow to reach an improved sensitivity on the search of Majorana neutrinos, achieving $T_{1/2}^{0\nu} \gtrsim 10^{26}$\,yr. 


%
%
%

\end{document}